\documentstyle[12pt,twoside,fleqn,espcrc1,epsfig,times]{article}

\title{The PHENIX Experiment at RHIC \\
\smallskip
{\normalsize D.P.~Morrison$^{4}$ for the PHENIX Collaboration}
}

\author{\scriptsize
Y.~Akiba,$^{16}$ 
O.~Alford,$^{20}$ 
M.~Allen,$^{30}$ 
W.~Allen,$^{30}$ 
G.~Alley,$^{30}$ 
Y.~Arai,$^{16}$ 
J.B.~Archuleta,$^{21}$ 
J.R.~Archuleta,$^{21}$ 
S.H.~Aronson,$^{4}$ 
I.~Atatekin,$^{24}$
% ...
\address{\scriptsize Collaborating Institutions:\\
% ...
$^{4}$ Brookhaven National Laboratory, Upton, NY 11973-5000, USA         \\
% ...
$^{16}$ KEK, High Energy Accelerator Research Organization, Tsukuba-shi, Ibaraki-ken 305, JAPAN      \\
% ...
$^{20}$ Lawrence Livermore National Laboratory, Livermore, CA 94550, USA        \\
$^{21}$ Los Alamos National Laboratory, Los Alamos, NM 87545, USA       \\
% ...
$^{24}$ McGill University, Montreal, Quebec H3A 2T8, CANADA         \\
% ...
$^{30}$ Oak Ridge National Laboratory, Oak Ridge, TN 37831, USA       \\
% ...
}}

\begin{document}

\maketitle

\begin{abstract}
  The physics emphases of the PHENIX collaboration and the design and
  current status of the PHENIX detector are discussed.  The plan of
  the collaboration for making the most effective use of the available
  luminosity in the first years of RHIC operation is also
  presented.\footnote{Visit \emph{http://www.rhic.bnl.gov/phenix} for
  the most current PHENIX information.}
\end{abstract}

\section{Physics and Design Aims}
\label{sec:physics}

The primary goals of the heavy-ion program of the PHENIX collaboration
are the detection of the quark-gluon plasma and the subsequent
characterization of its physical properties.  To address these aims,
PHENIX will pursue a wide range of high energy heavy-ion physics
topics.  The breadth of the physics program represents the expectation
that it will require the synthesis of a number of measurements to
investigate the physics of the quark-gluon plasma.  The broad physics
agenda of the collaboration is also reflected in the design of the
PHENIX detector itself, which is capable of measuring hadrons, leptons
and photons with excellent momentum and energy resolution.  PHENIX has
chosen to instrument a selective acceptance with multiple detector
technologies to provide very discriminating particle identification
abilities.  Additionally, PHENIX will take advantage of RHIC's
capability to collide beams of polarized protons with a vigorous spin
physics program, a subject covered in a separate contribution to these
proceedings\cite{Saito98}.

The first measurements PHENIX will make will be of global event
properties such as charged particle multiplicity, $E_T$ production,
the $\left< p_\perp \right>$ of charged particles, and fluctuations in
these quantities.  Charged particle multiplicity and $E_T$, alone or
in correlation with zero-degree calorimetry, will provide information
about the geometry of each collision. From these data one can also
deduce the energy density achieved in each event.  The geometry of the
collision, charge particle multiplicity, $E_T$ and energy density may
all be used to classify events for other analyses.

PHENIX will study many proposed signatures of the deconfinement
transition and the restoration of chiral symmetry.  The first of
these, the deconfinement transition, should produce a number of
signals observable in the PHENIX detector.  For instance, the
suppression of J/$\psi$ and $\psi'$ production relative to that of the
$\Upsilon$ will yield information about the strength of Debye
screening in the deconfined plasma.  Measuring J/$\psi$ suppression
relative to the Drell-Yan continuum will allow comparisons with
current results such as those from NA50\cite{NA50}.  Comparison of
charmonium production relative to that of open charm---primarily
identified through the semi-leptonic decay of charm mesons---will
allow PHENIX to disentangle initial state effects such as gluon
shadowing from the later dissolving of any created charmonium.  The
many $\textrm{D}\bar{\textrm{D}}$ pairs that are expected in central
Au+Au collisions will also give PHENIX a solid base from which to 
investigate open charm enhancement in the quark-gluon plasma.

An examination of chiral symmetry restoration will complement the
study of deconfinement.  The in-medium modification of meson
properties due to the restoration of chiral symmetry is predicted to
cause changes in the mass and width of the $\phi$ meson.  Since the
mass of the $\phi$ meson is only 33 MeV greater than twice the charged
kaon mass, changes in its properties will also affect the relative
branching ratio of $\phi$ mesons decaying via K$^+$K$^-$ or $e^+e^-$
channels.

The thermal history and available degrees of freedom will be studied
through direct $\gamma$ production and $\gamma^* \rightarrow e^+e^-,
\mu^+\mu^-$ channels.  Photons, like leptons, are unperturbed by the
strong interactions that plague hadronic signals and thus retain
information about the early history of the collision.  Whether the
colliding system forms a plasma with many degrees of freedom, remains
a hot hadronic gas, or evolves through a long-lived mixed state, all
have effects on the spectrum of emitted photons.  Very high $p_\perp$
photons may also serve as a reliable flag for an oppositely directed
jet, the properties of which may be measured via the leading particle
spectrum.

The measurement of bosonic or fermionic Hanbury-Brown Twiss
correlations and the coalescence likelihood of various nuclei and
anti-nuclei will give insights into the space-time extent and evolution
of heavy-ion collisions at RHIC.

Enhanced strangeness, already a staple feature of relativistic
heavy-ion physics, will be studied in PHENIX by determining the
production cross section of K$^\pm$ and $\phi$ mesons.  This will be
complemented by an investigation of enhanced charm production.

\section{Construction and Current Status of the Experiment}
\label{sec:status}

Fundamentally, the PHENIX detector consists of a large acceptance
charged particle detector and of four spectrometer arms---a pair of
spectrometers measuring electrons, photons and hadrons which straddles
mid-rapidity, and a pair of muon spectrometers at forward
rapidities---all working together in an integrated
manner\cite{PhenixCDR}.  Each of the four arms has a geometric
acceptance of approximately one steradian.  The magnetic field in the
volume of the collision region is axial, while the magnets of the muon
arms produce radial fields. The PHENIX detector is comprised of eleven
different subsystems, so that the task of integrating and
commissioning the detector is one of the biggest hurdles facing the
collaboration.

\begin{figure}[hbtp]
  \begin{center}
    \leavevmode
    \epsfig{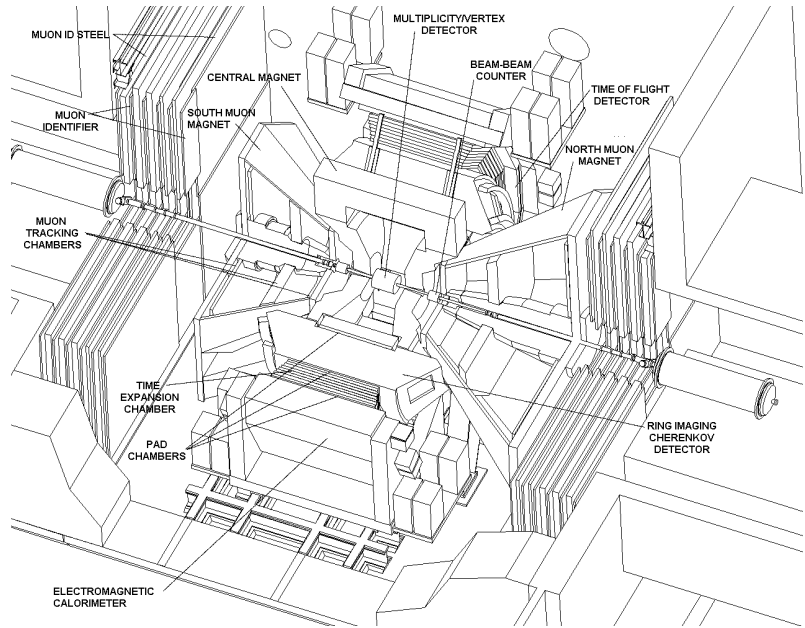}
    \caption{A cutaway drawing of the PHENIX experiment.  Labeled arrows
      indicate the major subsystems of the detector.}
    \label{fig:detector}
  \end{center}
\end{figure}

The main sources of event characterization information are the
beam-beam counter, which consists of two arrays of quartz \v{C}erenkov
telescopes surrounding the beam, and the multiplicity and vertex
detector, composed of concentric barrels of silicon strip detectors
and end-caps made of silicon pads.

Electromagnetic calorimeters are mounted outermost on each of the two
central arms.  PHENIX uses two technologies for calorimetry:
lead-scintillator with good timing properties, and lead-glass with
better energy resolution.

The central arm tracking system in PHENIX uses the information
provided by several detectors.  Pad chambers yield the
three-dimensional space points that are essential for pattern
recognition, drift chambers provide precise projective measurements of
particle trajectories, and time-expansion chambers provide $r$-$\phi$
information as well as particle identification.  Using this
tracking information the mass resolution of $\phi \rightarrow e^+e^-$
is determined to better than 0.5\% for $p_\perp < 2$ GeV/c.

Particle identification also hinges on several detectors.  Panels of
time of flight scintillators cover part of the central arm acceptance,
and the 85 ps timing resolution of this time of flight system
separates kaons from pions up to 2.5 GeV/c.  The timing resolution of
the lead-scintillator, 280 ps, can separate kaons from pions up to
about 1.4 GeV/c, and its large acceptance greatly improves the rates
for measurements such as $\phi\rightarrow \textrm{K}^+\textrm{K}^-$.
For electron identification, information from the ring-imaging
\v{C}erenkov detector, the $dE/dx$ measurement of the time-expansion
chamber, and information from the electromagnetic calorimeter are
combined to reject pion contamination of the identified electrons to
one part in $10^4$ over a wide range in momentum.

The first part of each muon arm (following a thick hadron absorber)
contains three stations of cathode strip tracking chambers.  The back
part of each arm consists of panels of Iarocci streamer tubes
alternating with plates of steel absorber.  The pion contamination of
identified muons is below one part in $10^4$, matching the high degree
of confidence in particle identification as is the case with the
central arm electron identification.  The excellent momentum
resolution of identified tracks in the muon arms yields a mass
resolution of 100 MeV/c$^2$ for J/$\psi \rightarrow \mu^+\mu^-$.

\section{Physics Opportunities Grow with Luminosity}
\label{sec:plans}

During its first two years of operation the luminosity of the RHIC
accelerator will gradually ramp up to its full design value.  In order
to make the most effective use of the available luminosity, the
collaboration has developed a plan which matches priorities for
physics studies to the anticipated profile of integrated luminosity.
Early in the first year of RHIC operation, when the luminosity will be
about 1\% of the design value, PHENIX will concentrate on measurements
such as $dN_{ch}/d\eta$, $dE_{T}/d\eta$, hadronic spectra, HBT and
inclusive $\gamma$ and $\pi^0$.  Each of these measurements can be
made with just a few $\mu \textrm{b}^{-1}$.  Toward the end of the
first year of operation, as the luminosity rises to 10\% of the design
value, measurements of $\phi\rightarrow \textrm{K}^+\textrm{K}^-$,
single high $p_T$ leptons and $J/\psi\rightarrow\mu^+\mu^-$ become
feasible.  By the end of the first year of RHIC operation, PHENIX
should have seen an integrated luminosity of roughly 100 $\mu
\textrm{b}^{-1}$.  It is in the second year of RHIC operation, as the
luminosity reaches its design goal, that the full physics program
becomes accessible.  At that point, the machine will have sufficient
luminosity for measurements of the Drell-Yan continuum, open charm
production $\Upsilon\rightarrow\mu^+\mu^-$, and J/$\psi$ and other
vector meson decays to $e^+e^-$.  The PHENIX spin program also becomes
possible in the second year of operation.  However, even this
luminosity does not exhaust the PHENIX appetite for physics.  As the
RHIC luminosity improves, the horizons of the PHENIX physics program
broaden still further.

\section{Acknowledgements}
\label{sec:acknowledgements}

 	We thank the technical staffs of the participating institutions 
for their vital contributions.  This detector construction project is
supported by the Department of Energy (U.S.A.), Monbu-sho and STA (Japan), 
RAS, RMAE, and RMS (Russia), BMBF (Germany), FRN and the Knut \& Alice 
Wallenberg Foundation (Sweden), and MIST and NSERC (Canada).


\begin{thebibliography}{9}
\bibitem{Saito98} N.~Saito, The PHENIX Spin Program.  These proceedings.
\bibitem{NA50} L. Ramello, NA38/50 Report.  These proceedings.
\bibitem{PhenixCDR} PHENIX Conceptual Design Report, BNL 1993
  (unpublished). 
\end{thebibliography}
\end{document}